\documentclass[twocolumn,showpacs,preprintnumbers,aps,prb,amsmath,amssymb]{revtex4}

\usepackage{graphicx}
\usepackage{dcolumn}
\usepackage{bm}

\begin{document}

\newcommand{\con}{cond-mat/}
\newcommand{\eli}{$\acute{{\rm E}}$liashberg }
\newcommand{\vi}{\vec{r}}
\newcommand{\vj}{\vec{r'}}
\renewcommand{\k}{\vec{k}}
\newcommand{\kk}{\vec{k'}}
\newcommand{\q}{\vec{q}}
\newcommand{\Q}{\vec{Q}}
\newcommand{\e}{\varepsilon}
\newcommand{\ee}{\varepsilon^{'}}
\newcommand{\s}{{\mit{\it \Sigma}}}
\newcommand{\J}{\mbox{\boldmath$J$}}
\newcommand{\vv}{\mbox{\boldmath$v$}}
\newcommand{\Jh}{J_{{\rm H}}}
\newcommand{\LL}{\mbox{\boldmath$L$}}
\renewcommand{\SS}{\mbox{\boldmath$S$}}
\newcommand{\Tc}{$T_{\rm c}$ }
\newcommand{\Tcf}{$T_{\rm c}$}
\newcommand{\etal}{{\it et al.} }
\newcommand{\PRL}{Phys. Rev. Lett. } 
\newcommand{\PRB}{Phys. Rev. B } 
\newcommand{\JPSJ}{J. Phys. Soc. Jpn. } 
\renewcommand{\vr}{\vec{r}}
\newcommand{\vrr}{\vec{r'}}
\newcommand{\vrrr}{\vec{r}_{2}}
\newcommand{\vrrrr}{\vec{r}_{3}}


\title{Angular Fulde-Ferrell-Larkin-Ovchinnikov state in cold fermion gases in a toroidal trap 
}

\author{Youichi Yanase$^{1,2}$}
\email{yanase@phys.sc.niigata-u.ac.jp}
\affiliation{%
$^{1}$ Department of Physics, University of Tokyo, Tokyo 113-0033, Japan 
\\
$^{2}$ Department of Physics, Niigata University, Niigata 950-2181, Japan 
}%

\date{February 4, 2009}

\begin{abstract}
 We study the angular Fulde-Ferrell-Larkin-Ovchinnikov 
(FFLO) state, in which the rotation symmetry is spontaneously broken, 
in population imbalanced fermion gases near the BCS-BEC crossover. 
 We investigate the superfluid gases at low temperatures 
on the basis of the Bogoliubov-de Gennes equation, 
and examine the stability against thermal fluctuations 
using the $T$-matrix approach beyond the local-density approximation. 
 We find that the angular FFLO state is stabilized in the gases 
confined in the toroidal trap but not in the harmonic trap. 
 The angular FFLO state is stable near the BCS-BEC crossover 
owing to the formation of pseudogap. 
 Spatial dependences of number density and local population imbalance 
are shown for an experimental test. 
\end{abstract}

\maketitle

 Superfluidity in cold fermion gases provides vast opportunities to 
study novel quantum phenomena.\cite{giorginireview} 
 One of the goals of current studies is the realization of 
the Fulde-Ferrell-Larkin-Ovchinnikov (FFLO) state \cite{FF,LO} in 
population imbalanced superfluid gases.\cite{Partridge01272006,Zwierlein01272006}
 FFLO superfluidity/superconductivity is attracting growing interest 
in a variety of fields, such as condensed-matter physics,\cite{matsuda2007} 
astrophysics, and nuclear physics.\cite{casalbuoni2004} 
 Since many parameters can be experimentally controlled,\cite{giorginireview} 
cold fermion gases are promising candidates for the FFLO state.\cite{sheehy:060401,yoshida:063601}

 Spontaneous breaking of space symmetry is a characteristic feature 
of the FFLO state. However, no firm  evidence has been obtained 
for the space symmetry breaking in the condensed-matter physics. 
 In contrast to the superconductors, the spatial structure of superfluid 
is directly measured in cold atom gases. Therefore, it is highly desired to 
observe the space symmetry breaking due to the FFLO superfluidity in 
cold fermion gases. 
 While the translation symmetry plays a major role in superconductors, 
cold atom gases lack translation symmetry owing to the trap potential. 
 Instead, the rotation symmetry is well defined in the latter.  
 The purpose of this Rapid Communication is to investigate the FFLO state 
which spontaneously breaks the rotation symmetry.

 On the basis of the mean-field Bogoliubov-de Gennes (BdG) equations, 
some authors investigated the radial FFLO (R-FFLO) state\cite{PhysRevA.72.025601,kinnunen:110403,machida2006gpd,mizushima2007,liu:023614} 
in which the order parameter changes its sign 
along the radial direction around the edge of the harmonic trap. 
 However, no space symmetry is broken in the R-FFLO state, 
and therefore it is difficult to distinguish it from the phase separated 
state.\cite{yi:031604,haque:011602,parish2007ftp}  
 In this Rapid Communication we show that the angular FFLO (A-FFLO) state 
with broken rotation symmetry is stabilized 
in the toroidal trap.\cite{ryu:260401} 
Several experiments are proposed for an unambiguous evidence 
for the A-FFLO state. 

 The superfluidity has been realized in the imbalanced fermion gases 
near the BCS-BEC crossover.\cite{Partridge01272006,partridge:190407,Zwierlein01272006,2006Natur.442...54Z,shin:030401}   
 However, that is not achieved in the BCS limit 
since the transition temperature \Tc is too small. 
 Because the mean-field theory breaks down near the BCS-BEC 
crossover,\cite{Chen20051,giorginireview} a theoretical treatment 
beyond the BdG equations is desired for the study of cold fermion gases. 
 To this end, the local-density approximation (LDA) has been used 
in the literature.\cite{sheehy:060401,yi:031604,haque:011602,parish2007ftp} 
 However, a theory beyond the LDA is needed to study the 
superfluid phase with broken space symmetry. 
 For these theoretical requirements we adopt the real-space 
self-consistent $T$-matrix approximation (RSTA).\cite{yanase2006,yanase-2008diamond}
 The reliability of the RSTA has been examined in the uniform system 
by comparing it with the nonperturbative infinite-loop order 
theory.\cite{yanase2004infinite} 
 We found that the RSTA is quantitatively valid at least 
in the BCS side of BCS-BEC crossover.

 We here investigate the gases confined 
in the (quasi-)two-dimensional space. 
 The two-dimensional gas is produced in the pancake 
potential $\omega_{\rm z} \gg \omega_{\perp}$ with  
$\omega_{\rm z}$ and $\omega_{\perp}$ being the harmonic trap frequency 
along the axial and radial directions, respectively. 
 The one-dimensional optical lattice along the axial direction also 
produces the quasi-two-dimensional gas.\cite{Hadzibabic2006}  
 Since the fluctuation completely suppresses the continuous symmetry breaking 
in one- and two-dimensional systems at finite temperatures, 
a weak three dimensionality is assumed to realize the state with 
broken space symmetry. 
 The following calculation is carried out in the two-dimensional model 
for simplicity, and the singularity of the low-dimensional model is 
cut off by a phenomenological procedure.

 We adopt the lattice Hamiltonian given as 
\begin{eqnarray}
\label{eq:attractive-Hubbard-model}
&& \hspace*{-10mm}
H= -t \sum_{<\vi,\vj>,\sigma} c_{\vi,\sigma}^{\dag}c_{\vj,\sigma} 
+ \sum_{\vi \sigma} (V(|\vi-\vi_{0}|) - \mu_{\sigma}) \hspace{0.5mm} n_{\vi,\sigma}
\nonumber \\ && \hspace*{0mm}
+ U \sum_{\vi} n_{\vi,1} \hspace{0.5mm} n_{\vi,2}, 
\end{eqnarray}
where $\sigma = 1,2$ denote two hyperfine states, 
$\vi_{0}$ is the center of the trap, and 
$n_{\vi,\sigma} = c_{\vi,\sigma}^{\dag}c_{\vi,\sigma}$ 
is the number operator of $\sigma$ particles. 
 We take the unit $\hbar = c = 1$. 
 The symbol $<\vi,\vj>$ denotes the summation over nearest neighbor sites. 
 The chemical potential $\mu_{\sigma}$ for $\sigma$ particles 
is determined so that the number of each particle is $N_{\sigma}$.  
 The particle number and the imbalance are expressed as 
$N = N_{1} + N_{2}$ and $P=(N_{1} - N_{2})/(N_{1} + N_{2})$, respectively. 
 The lattice model is adopted for simplicity, but the discreteness 
of the lattice is negligible since we assume a small particle density 
$N/N_{\rm L} = 0.1$, 
where $N_{\rm L} = L \times L$ is the number of lattice sites. 
 Therefore, the following results are valid for continuous systems 
without lattices in the two-dimensional space. 
 Since $N_{\rm L} = 38 \times 38$ in our calculation, 
the particle number is $N \sim 144$. 
 We take the unit of length  $d$ so that $1/2 m d^{2}=t=1$, where 
$m$ is the mass of atoms. 
 We define the Fermi energy as $\e_{\rm F} = \mu - \e_{0}$, where 
$\e_{0}$ is the energy of the lowest eigenstate and $\mu$ is the chemical 
potential at $P=0$ and $U=0$. 
 We find that the superfluidity is the leading instability 
and no spin/charge density wave occurs in this model.

 The last term of Eq.~(1) describes the $s$-wave attractive interaction. 
 We assume $U/t=-5$, which leads to $U/\e_{\rm F}=-3.1$. 
 The BCS-BEC crossover is characterized in the two-dimensional system 
through the two-particle binding energy $e_{\rm b}$ rather than 
the three-dimensional scattering length $a_{\rm s}$.\cite{randeria2d} 
 In the uniform system, the binding energy is related with the 
chemical potential shift as 
$\Delta \mu = \mu - \mu_{0} = - \frac{e_{\rm b}}{2} $, where $\mu_{0}$ is the 
chemical potential shifted by the Hartree term. 
 The order parameter is described as $\Delta_{0} = \sqrt{2 e_{\rm b} \e_{\rm F}}$. 
 Our calculations of $\Delta \mu$ and $\Delta_{0}$ consistently lead to 
$ e_{\rm b} \sim 0.43 \e_{\rm F}$ for $U/t=-5$. 
Since the BCS-BEC crossover occurs around $ e_{\rm b} \sim \e_{\rm F}$,  
our model is close to the BCS-BEC crossover slightly in the BCS side. 
 The binding energy $e_{\rm b}$ and effective interaction $U$ are 
related with the three-dimensional scattering length  $a_{\rm s}$ 
through the confinement length 
$a_{\rm z} = 1/\sqrt{m \omega_{\rm z}}$.\cite{kestner:063610}

 The trap potential is assumed to be 
$V(r)= \frac{1}{2} \omega_{\rm ho} (r/r_{0})^{2} 
+ \omega_{\rm tr} \exp(-r/\xi) $. 
 This potential describes the harmonic trap for $\omega_{\rm tr}=0$ 
and the toroidal trap for $\omega_{\rm tr} \ne 0$. 
 We found that the A-FFLO state is unstable in the harmonic trap 
in the whole parameter range. 
 Therefore, we here show the results for the toroidally trapped system with 
$\omega_{\rm ho} = 12$, $\omega_{\rm tr} = 8$, and $\xi=5$. 
 The A-FFLO state is stabilized for any $\omega_{\rm tr}/\omega_{\rm ho} > 0$, 
whose reason will be discussed later.

 We first analyze the model within the mean-field BdG equation 
and later investigate the role of thermal fluctuations 
on the basis of the RSTA. 
 We obtain the mean-field Hamiltonian of BdG equations as 
%
$  H = -t \sum_{<\vi,\vj>,\sigma} c_{\vi,\sigma}^{\dag}c_{\vj,\sigma}
  + \sum_{\vi,\sigma} W_{\sigma}(\vi) n_{\vi,\sigma}, 
  - \sum_{\vi} [\Delta(\vi) c_{\vi,1}^{\dag} c_{\vi,2}^{\dag}
  + c.c.]$, 
%
where 
$W_{\sigma}(\vi) = V(|\vi-\vi_{0}|) + U n_{\bar{\sigma}}(\vr) - \mu_{\sigma}$, 
$\bar{\sigma}= 3-\sigma$, $n_{\sigma}(\vr) = <n_{\vi,\sigma}>$, and 
$\Delta(\vi) = U <c_{\vi,1} c_{\vi,2}>$. 
 The unphysical ultra-violet divergence in $\Delta(\vi)$ 
\cite{giorginireview,randeria2d} is naturally cut off 
since we adopt the lattice model. 
 We numerically determine the stable phase by comparing the free energy 
of self-consistent solutions for $n_{\sigma}(\vr)$ and $\Delta(\vi)$.

 The RSTA has been formulated for the inhomogeneous 
superconductors.\cite{yanase2006,yanase-2008diamond} 
 The Green's function $G^{\sigma}(\vr,\vrr,\omega_{n})$ 
and the $T$-matrix $T(\vr,\vrr)$ are obtained by 
the following self-consistent equations: 
\begin{eqnarray}
\label{eq:G}
&& \hspace*{-5mm}  
G^{\sigma}(\vr,\vrr,\omega_{n})= 
G_{0}^{\sigma}(\vr,\vrr,\omega_{n}) + 
\nonumber \\ && \hspace*{0mm}
\sum_{\vrrr,\vrrrr}
G_{0}^{\sigma}(\vr,\vrrr,\omega_{n})
\Sigma^{\sigma}(\vrrr,\vrrrr,\omega_{n})
G^{\sigma}(\vrrrr,\vrr,\omega_{n}), 
\\ && \hspace*{-5mm} 
\label{eq:Sigma}
\Sigma^{\sigma}(\vr,\vrr,\omega_{n}) = 
U n_{\bar{\sigma}}(\vr) \delta_{\vr,\vrr}  - 
T U^{2} T(\vr,\vrr) G^{\bar{\sigma}}(\vrr,\vr,-\omega_{n}), 
\nonumber \\
\\ && \hspace*{-5mm} 
\label{eq:T}
T(\vr,\vrr) = T_{0}(\vr,\vrr) - 
\sum_{\vrrr}
U T_{0}(\vr,\vrrr)  
T(\vrrr,\vrr), 
\\ &&  \hspace*{-5mm} 
\label{eq:T0}
T_{0}(\vr,\vrr)
=T \sum_{n} G^{1}(\vr,\vrr,\omega_{n}) G^{2}(\vr,\vrr,-\omega_{n}), 
\end{eqnarray} 
where $G_{0}^{\sigma}(\vr,\vrr,\omega_{n})$ is the Green's function for $U=0$, 
$\omega_{n} = (2 n + 1) \pi T$ is the Matsubara frequency, 
and $T$ is the temperature. 
 The thermal fluctuation, which is neglected in the BdG equations, 
is taken into account in the self-energy $\Sigma^{\sigma}(\vr,\vrr,\omega_{n})$ 
in the one-loop order.  
 The pseudogap in the single-particle excitation as well as the 
shift in chemical potential are taken into account in the RSTA. 
 The former is neglected in the often-used Nozieres and 
Schmitt-Rink theory.\cite{NSR} 
 We show that the pseudogap plays an essential role for the stability of 
A-FFLO state.

 The quantum fluctuation is ignored in the RSTA, and therefore the RSTA 
is valid at finite temperature around \Tcf.\cite{yanase2004infinite} 
 Therefore, the RSTA is used to determine the instability to 
the superfluid state. 
 The \Tc is determined by the Thouless criterion. 
The maximum eigenvalue of $|U| T_{0}(\vr,\vrr)$, namely, 
$\lambda_{\rm L}$, is unity at $T=T_{\rm c}$. 
 Since the true long-range order does not occur in finite systems, 
we adopt the criterion $\lambda_{\rm L} = 1 - \delta_{\rm L} = 0.98$ for 
$T_{\rm c}$ below which the long-range coherence develops. 
 The singularity due to the one or two dimensionalities is also cut off 
by this procedure. 
 The following results are not qualitatively altered by the choice of 
$\delta_{\rm L}$. This means that the following results are not sensitive 
to the three dimensionality which is phenomenologically 
taken into account by a finite $\delta_{\rm L}$. 
 The density of states (DOS) for particles $\sigma$ is obtained as  
$\rho^{\sigma}(\omega) = 
- \frac{1}{\pi N_{\rm L}}\sum_{\vr} {\rm Im}G^{\sigma {\rm R}}(\vr,\vr,\omega)$,  
where $G^{\sigma {\rm R}}(\vr,\vr,\omega)$ is the retarded Green's function. 
 The total DOS is expressed as 
$\rho(\omega) = \rho^{1}(\omega) + \rho^{2}(\omega)$.

\begin{figure}
\includegraphics[width=6.5cm]{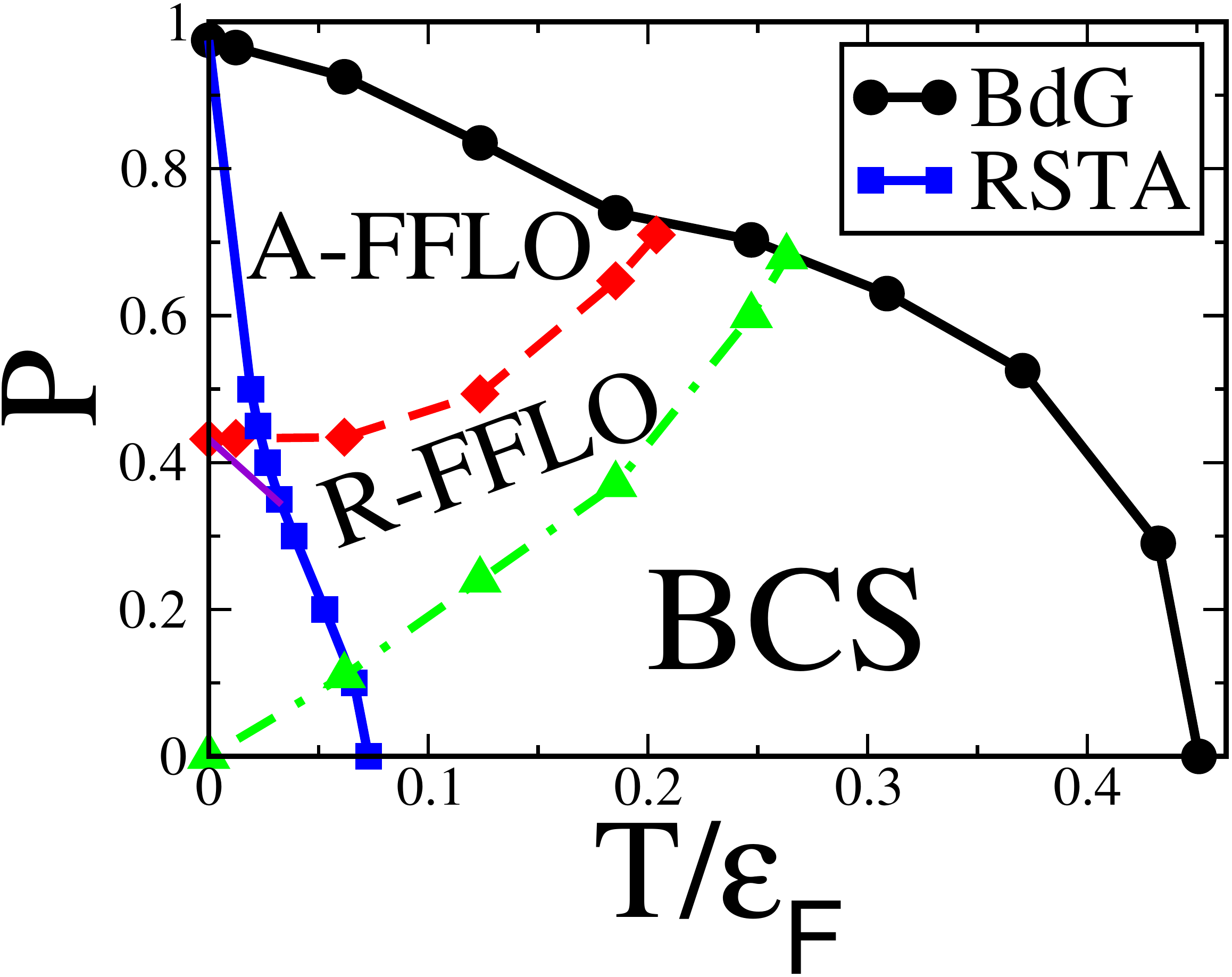}
\caption{\label{fig:phasediagram} 
Phase diagram for the imbalance $P$ and the reduced temperature 
$T/\e_{\rm F}$. 
 Phase boundaries obtained by the BdG equations are shown 
by circles, triangles, and diamonds. 
 The BCS state, R-FFLO state, and A-FFLO state are shown in the figure. 
 The phase diagram determined by the RSTA is shown by the squares 
and (purple) thin solid line. 
 The A-FFLO state is stable above the thin solid line, 
while the R-FFLO or BCS state is stable below it. 
We fix the particle density $N/N_{\rm L} = 0.1$ in all figures. 
}
\end{figure}

 We first discuss the results of BdG equations. 
 Although the quantitatively reliable result is not obtained 
near the BCS-BEC crossover, the properties of each phase, 
such as the local population imbalance, are captured by the BdG equations. 
 We here study the spatial structure of several superfluid phases. 
 Figure~1 shows the phase diagram in which 
BCS state, R-FFLO state, and A-FFLO state are stabilized. 
 Since the BCS state smoothly changes to the R-FFLO state 
without any phase transition, 
we show the crossover line above which the superfluid order parameter 
changes its sign around the trap edge. 
 With an increase in the population imbalance, the second-order phase 
transition occurs from the R-FFLO state to the A-FFLO state. 
 Although the transition temperature is significantly overestimated 
in BdG equations, the successive phase transitions from 
the BCS state to the A-FFLO state are not altered by the fluctuations, 
as will be shown on the basis of the RSTA.

\begin{figure}
\hspace*{-10mm}
\includegraphics[width=8cm]{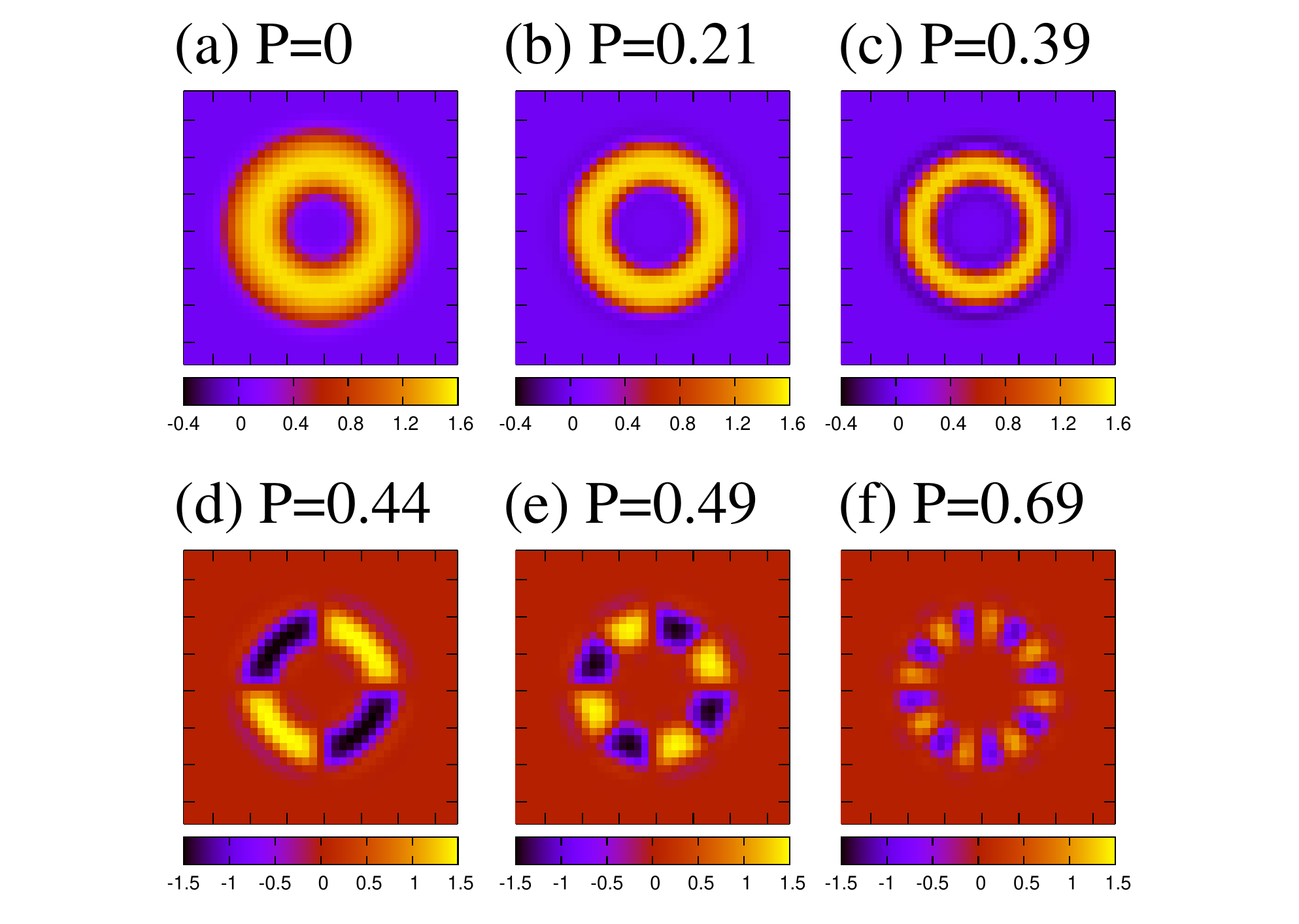}
\caption{
Spatial dependence of superfluid order parameter $\Delta(\vr)$ 
at $T/\e_{\rm F}=0.00062$. 
(a) $P=0$, (b) $0.21$, (c) $0.39$, (d) $0.44$, (e) $0.49$, 
and (f) $0.69$, respectively. 
}
\end{figure}

 We clarify the spatial structures of each phase in Figs.~2-4. 
 First we show the order parameter for various  
imbalances in Fig.~2. 
 Figure~2(a) shows the BCS state realized in the balanced gas. 
 Figures~2(b) and 2(c) show the R-FFLO state in the imbalanced gases. 
 The rotation symmetry is preserved in these states. 
 As the imbalance is increased in these states, 
the width of superfluid region shrinks. 
 This deformation is regarded as the {\it self-one-dimensionalization} 
of the superfluid along the angular direction and leads to 
the A-FFLO state for $P > 0.43$. 
 Figures~2(d)-2(f) show the spontaneous rotation symmetry breaking 
in the A-FFLO state. 
 Thus, the A-FFLO state is an analog of the (quasi-)one-dimensional 
FFLO state.\cite{PhysRevB.30.122,orso:070402,hu:070403,
parish:250403,feiguin:220508,tezuka:110403,Ye:JPhys2009}
 It is known that the FFLO state is favored in the quasi-one-dimensional 
system because of the nesting of Fermi surface.\cite{matsuda2007} 
 An important finding of this Rapid Communication is the spontaneous formation of 
quasi-one-dimensional superfluid in the toroidal trap without any fine tuning. 
 This should be contrasted to the harmonic trap, in which 
the uniform one-dimensional superfluid is hardly produced. 
 We found that the self-one-dimensionalization occurs and the A-FFLO state 
is stabilized for any value of $\omega_{\rm tr}/\omega_{\rm ho} > 0$. 
 As the particle density decreases, the quasi-one-dimensional structure 
is enhanced, and therefore the A-FFLO state is favored. 
 A similar spatial structure has been discussed within the purely
one-dimensional model; however, the self-one-dimensionalization has not 
been noticed.~\cite{Ye:JPhys2009}  
 Although we investigate the gases in the (quasi-)two-dimensional trap 
for simplicity, the A-FFLO state will be stabilized by the 
self-one-dimensionalization in a more general three-dimensional trap too.

\begin{figure}
\hspace*{-10mm}
\includegraphics[width=8cm]{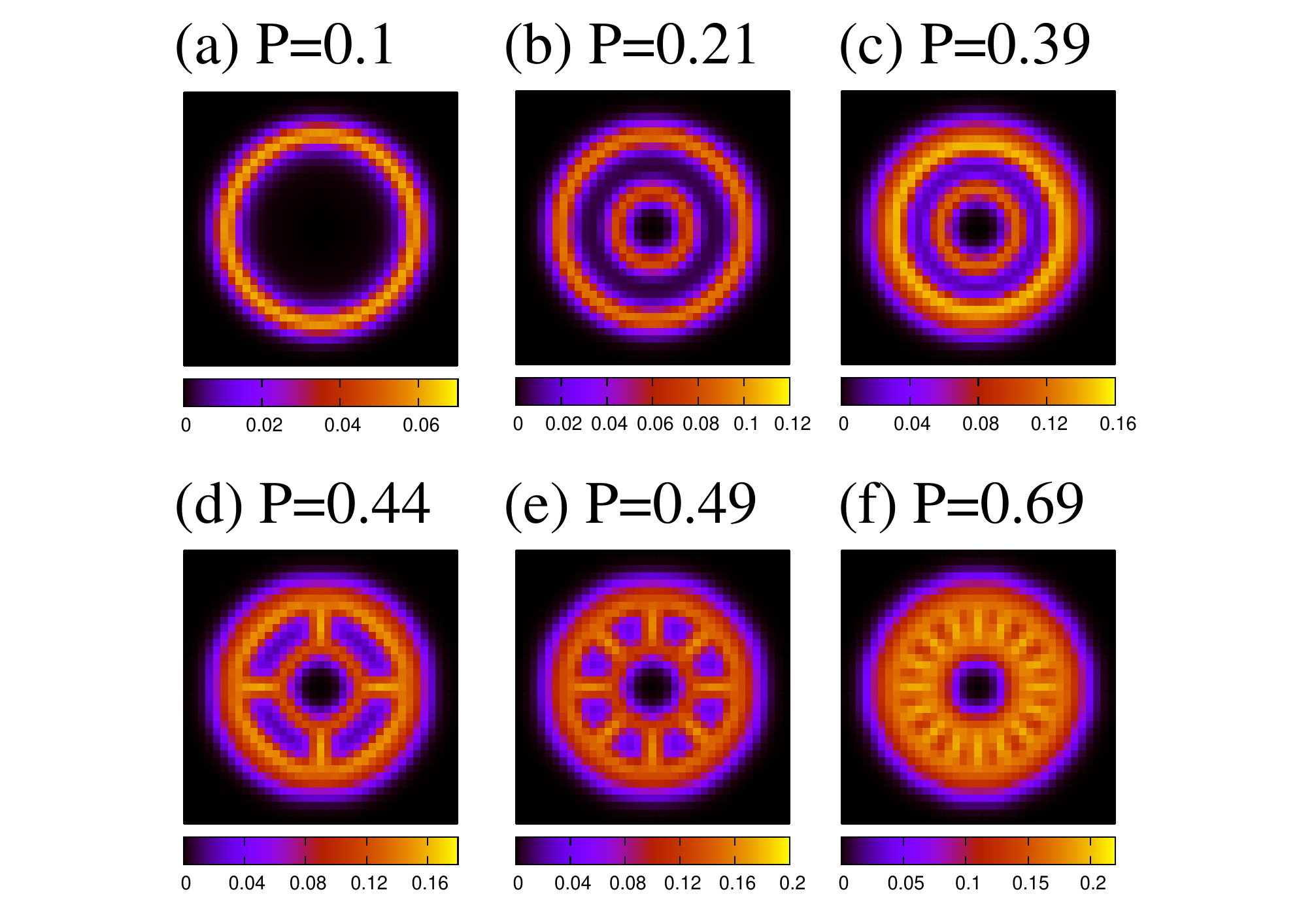}
\caption{
Spatial dependence of local population imbalance $n_{1}(\vr)-n_{2}(\vr)$. 
We assume $P=0.1$ in (a). The other parameters are 
the same as in Fig.~2.  
}
\end{figure}

 Figure~3 shows the spatial dependence of local population imbalance 
$n_{1}(\vr)-n_{2}(\vr)$. 
 While the population imbalance appears around the outer and/or inner edges 
in the R-FFLO state [Figs.~3(a)-3(c)], 
the spontaneous rotation symmetry breaking is clearly shown 
in the A-FFLO state [Figs.~3(d)-3(f)]. 
 A clear four-fold anisotropy is shown in Fig.~3(d), while 
the spatial dependence is smeared with an increase in the imbalance, 
as shown in Fig.~3(f). 
 Thus, the features of the A-FFLO state are pronounced near the 
phase boundary to the R-FFLO state. 
 A characteristic feature of A-FFLO state also appears 
in the particle density $n_{1}(\vr)+n_{2}(\vr)$ as shown in Figs.~4(d)-4(f). 
 The particle density decreases around the spatial nodes 
to gain the condensation energy. 
 Owing to the spontaneous symmetry breaking, many A-FFLO states with 
different nodal directions are essentially degenerate. 
This degeneracy is slightly lifted by the lattice in our calculation. 
We show the most stable states in Figs.~2-4.

\begin{figure}
\hspace*{-10mm}
\includegraphics[width=8cm]{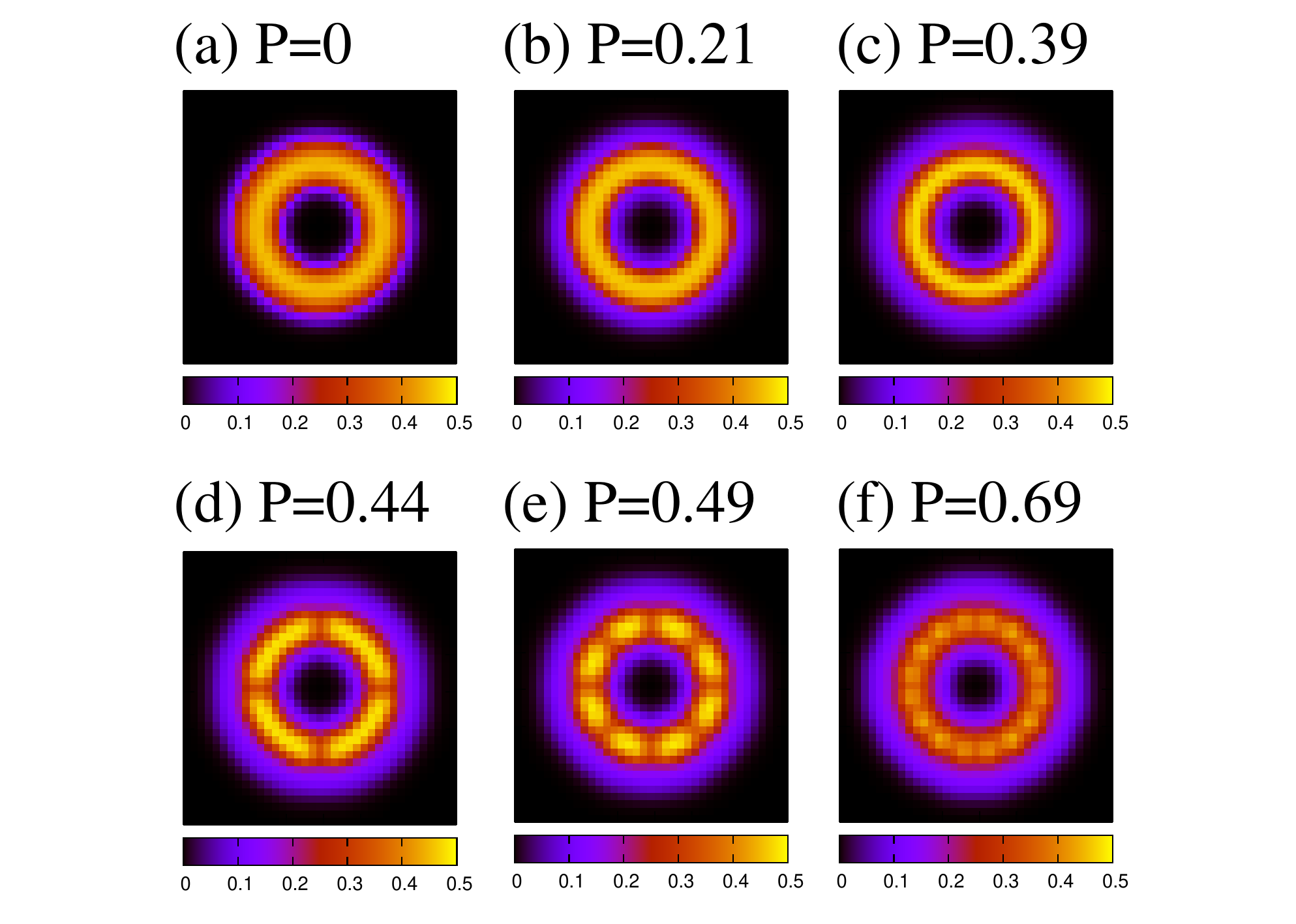}
\caption{
Spatial dependence of particle density $n_{1}(\vr)+n_{2}(\vr)$. 
The parameters are the same as in Fig.~2. 
}
\end{figure}

\begin{figure}
\hspace*{-5mm}
\includegraphics[width=8cm]{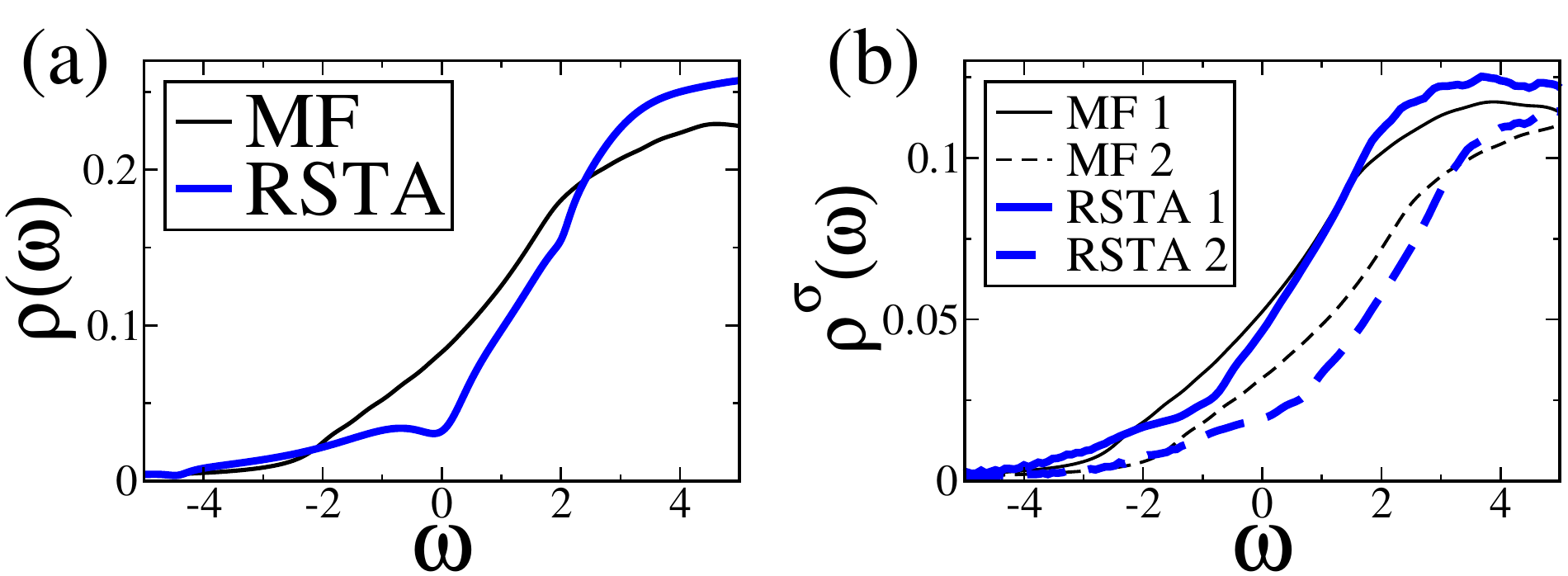}
\caption{
(a) Total DOS $\rho(\omega)$ in the balanced gas ($P=0$) at 
$T/\e_{\rm F}=0.074$. 
(b) Spin resolved DOS $\rho^{\sigma}(\omega)$ in the imbalanced gas 
($P=0.4$) at $T/\e_{\rm F}=0.028$. Solid and dashed lines show the DOS for 
$\sigma=1$ and $2$, respectively. 
 Thick lines show the results of RSTA, while thin lines 
are obtained by the mean-field theory for the normal fluid 
with $\Delta(\vr)=0$. 
}
\end{figure}

 We here turn to the results of RSTA, and discuss the roles of 
the thermal fluctuation. 
 The phase diagram is shown in Fig.~1 
and is compared to the mean-field BdG equations. 
 According to the RSTA, the phase transition to the A-FFLO state 
occurs above $P=0.35$, although the \Tc is decreased by the fluctuation.  
 This means that the A-FFLO state is stable against the fluctuation, 
in contrast to the previous studies.\cite{shimahara1998,ohashi2002}   
 We see the stability of the A-FFLO state because the phase diagram is 
plotted for the imbalance $P$, but not for the 
``magnetic field'' $\mu_{1}-\mu_{2}$. 
 The A-FFLO state seems to be suppressed by the fluctuation when the 
phase diagram is plotted for the ``magnetic field''  
as in the studies of superconductors.\cite{matsuda2007}  
 The relation between the ``magnetic field'' $\mu_{1}-\mu_{2}$ 
and the imbalance $P$ is affected by the pseudogap. 
 In Fig.~5 we see the decrease in DOS around $\omega=0$, namely, the pseudogap. 
 The ``magnetic field'' for a fixed imbalance is increased 
by the pseudogap since the ``spin susceptibility'' 
$\chi = P/(\mu_{1}-\mu_{2})$ decreases. 
 The large ``magnetic field'' leads to the large splitting of Fermi surfaces 
for particles $1$ and $2$ and stabilizes the A-FFLO state. 
 In other words, the A-FFLO state is stable near the BCS-BEC crossover 
in cold fermion gases since the spin-diffusion time is long enough to 
conserve the imbalance. 
 This should be contrasted to the superconductors in which the magnetization 
is not conserved.

 We here comment on the superfluid state in the BCS and BEC regimes. 
 In the BCS regime the phase diagram obtained by the BdG equation (Fig.~1) 
is reliable although the \Tc and Chandrasekhar-Clogston limit are decreased. 
 On the other hand, the imbalanced gas in the BEC limit is described 
by the mixture of molecular bosons and remaining fermions, and then, 
the FFLO state is not stabilized.

 In this Rapid Communication we focused on the superfluid 
with broken rotation symmetry. 
 Such a spontaneous symmetry breaking is not allowed in the purely one- or 
two-dimensional systems\cite{shimahara1998,ohashi2002}  
but is expected to be realized in the weakly three-dimensional system. 
 On the other hand, it is also interesting to investigate the gases 
in the two-dimensional toroidal trap. Then, the rotation 
symmetry breaking is suppressed by the gapless collective mode in the 
isotropic trap but is produced in the anisotropic trap 
with $\omega_{\rm x} \ne \omega_{\rm y}$. 
 Such a giant response to the trap anisotropy may manifest the 
tendency to the rotation symmetry breaking as well as the singularity of 
low-dimensional systems.

 In summary, we found that the A-FFLO state is stabilized in the population 
imbalanced fermion gases confined in the toroidal trap.  
 The formation of the R-FFLO state leads to the self-one-dimensionalization 
of the superfluid and stabilizes the A-FFLO state 
in the highly imbalanced gases. 
 Then, the rotation symmetry is spontaneously broken.
 The search for the FFLO state in cold fermion gases has been fruitless 
probably because the experiments were carried out for the harmonically 
trapped gases. 
 It is difficult to detect the FFLO state in the harmonic trap since 
no space symmetry breaking occurs. 
 We suggest that the experiment in the toroidal trap will realize 
the FFLO state with broken rotation symmetry 
and will obtain the unambiguous evidence for the FFLO state 
which has been searched for more than 40 years after the theoretical 
predictions.\cite{FF,LO}

 Recently we have become aware of the paper 
by Chen {\it et al.} in which a superfluid state similar to  
the A-FFLO state was investigated in the optical lattice.\cite{chen:054512}  
 However, the rotation symmetry is not well-defined in the optical lattice. 
 The superfluidity near the BCS-BEC crossover has not been investigated.

 We are grateful to T. Mizushima, M. Okumura, Y. Ohashi, 
M. Tezuka, S. Tsuchiya, and M. Ueda for fruitful discussions. 
 This study has been supported by 
Grants-in-Aid for Scientific Research 
from the MEXT (Grants No. 20029008, No. 20740187, and No. 21102506). 
 Numerical computation was carried out 
at the Yukawa Institute Computer Facility.

\bibliographystyle{apsrev}
\bibliography{toroidalfflo}

\end{document}